\documentclass[fleqn,usenatbib]{mnras}

\usepackage{newtxtext,newtxmath}

\usepackage[T1]{fontenc}
\usepackage{ae,aecompl}

\usepackage{subfig}

\usepackage{graphicx}   
\usepackage{amsmath}    
\usepackage{amssymb}    

\title[Modelling Pop. III chemical enrichment]{Modelling the chemical enrichment of Population III supernovae: The origin of the metals in near-pristine gas clouds.}

\author[L. Welsh et al.]{
Louise Welsh$^{1}$\thanks{E-mail: louise.a.welsh@durham.ac.uk}, Ryan Cooke$^{1}$, and Michele Fumagalli$^{1,2}$ \\
$^{1}$Centre for Extragalactic Astronomy, Durham University, South Road, Durham DH1 3LE, UK \\
$^{2}$Institute for Computational Cosmology, Durham University, South Road, Durham DH1 3LE, UK \\}

\date{Accepted 2019 May 29. Received 2019 May 22; in original form 2019 March 20}

\pubyear{2019}

\defcitealias{HegerWoosley2010}{HW10}
\defcitealias{LimongiChieffi2012}{LC12}
\defcitealias{WoosleyWeaver1995}{WW95}

\begin{document}

\label{firstpage}
\pagerange{\pageref{firstpage}--\pageref{lastpage}}
\maketitle

\begin{abstract}
The most metal-poor, high redshift damped Lyman $\alpha$ systems (DLAs) provide a window to study some of the first few generations of stars. In this paper, we present a novel model to investigate the chemical enrichment of the near-pristine DLA population. This model accounts for the mass distribution of the enriching stellar population, the typical explosion energy of their supernovae, and the average number of stars that contribute to the enrichment of these DLAs. We conduct a maximum likelihood analysis of these model parameters using the observed relative element abundances ([C/O], [Si/O], and [Fe/O]) of the 11 most metal-poor DLAs currently known. We find that the mass distribution of the stars that have enriched this sample of metal-poor DLAs can be well-described by a Salpeter-like IMF slope at $M>10~{\rm M}_{\odot}$ and that a typical metal-poor DLA has been enriched by $\lesssim 72$ massive stars (95 per cent confidence), with masses $\lesssim40~{\rm M_{\odot}}$. The inferred typical explosion energy ($\hat{E}_{\rm exp}=1.8^{+0.3}_{-0.2}\times10^{51}~{\rm erg}$) is somewhat lower than that found by recent works that model the enrichment of metal-poor halo stars. 
These constraints suggest that some of the metal-poor DLAs in our sample may have been enriched by Population II stars. Using our enrichment model, we also infer some of the typical physical properties of the most metal-poor DLAs. We estimate that the total stellar mass content is $\log_{10}(M_{\star}/{\rm M_{\odot}})=3.5^{+0.3}_{-0.4}$ and the total gas mass is $\log_{10}({M_{\rm gas}}/~{\rm M_{\odot}})= 7.0^{+0.3}_{-0.4}$ for systems with a relative oxygen abundance [O/H]~$\approx -3.0$.

\end{abstract}



\section{Introduction}
The first stars in the Universe necessarily formed out of a primordial environment, heralding an epoch known as the cosmic dawn, at a redshift of $z\sim 20- 30$ \citep{BarkanaLoeb2001}. At high densities, collapsing primordial gas relied chiefly on molecular hydrogen, an inefficient coolant, to radiate energy and facilitate collapse. As a result, it is believed that primordial gas was unable to form low mass stars; instead, small multiples of relatively massive stars are thought to have formed in small clusters \citep{Abel2002,Glover2013}. Elements heavier than lithium, known as metals, were forged within the cores of these first stars. When the first stars ended their lives, some as supernovae (SNe) explosions, the surrounding gas was enriched with these heavy elements, altering the process of all subsequent star formation. The incorporation of metals into star-forming gas facilitates numerous cooling pathways. Metal-enriched gas can therefore collapse and fragment more effectively than primordial gas.  The unique formation history of the first, metal-free, population is expected to be evident from its stellar Initial Mass Function (IMF) --- the characteristic mass of which is thought to be relatively larger than that of populations which form from metal-enriched gas. \\
 \indent  Lacking direct observations, the most direct means to pin down the mass distribution of metal-free stars is to simulate their formation in a cosmological setting (e.g. \citealt{Tegmark1997}; \citealt{BarkanaLoeb2001}; \citealt{Abel2002}; \citealt{BrommCoppiLarson2002};  \citealt{Turk2009}; \citealt{Greif2010}; \citealt{Clark2011}; \citealt{Hirano2014}; \citealt{Stacy2016}). Overall, these works indicate that the first stars, also known as Population III (or Pop III) stars, had masses in the range of $10 - 100$~M$_{\odot}$ and formed obeying a relatively bottom-light distribution compared with that of star formation today (see \citealt{Bastain2010} for a recent review). These massive stars would have had distinctly short lifetimes; none could have survived long enough to be observed today. The fact that a metal-free star has yet to be detected, in spite of both historic and on-going surveys (e.g. \citealt{Bond1980, Beers1985, Ryan1991, Beers1992, McWilliam1995, Ryan1996, Cayrel2004, Beers2008, Christlieb2008, Roederer2014, Howes2016, Starkenburg2017}), supports these theoretical works.  \\
 \indent We can observationally probe the properties of this potentially extinct population via indirect methods. Namely, we search for the unique chemical fingerprint that metal-free stars leave behind once they explode as Type II core-collapse SNe. To reliably infer the properties of Population III stars, we must therefore isolate systems that have only been chemically enriched by the SNe of metal-free stars. Historically, this has been achieved by searching for surviving Extremely Metal-Poor (EMP) stars, which are characterised by an iron abundance that is 1000 times less than that of the Sun\footnote{The use of iron as a metallicity tracer is a consequence of our ability to reliably detect its associated absorption features in stellar spectra.} (see \citealt{Beers2005} and \citealt{FrebelNorris2015} for a review of this field). These surviving EMP stars were among the second generation of stars to form in the Universe and may have been exclusively enriched by Population III SNe. \\ 
 \indent As suggested by \cite{Erni2006}, \cite{Pettini2008}, \cite{Penprase2010}, and \cite{Crighton2016}, it is also possible to search for the signatures of Population III stars in the large reservoirs of neutral hydrogen that are found along the line-of-sight towards unrelated, background quasars. 
 The relative metal abundances of these gaseous systems are encoded with information about the stars that have contributed to their enrichment. Thus, the most metal-deficient systems are invaluable tools for studying the earliest episodes of chemical enrichment. Indeed, some of the most metal-poor gaseous systems may have been solely enriched by the first generation of stars (e.g. \citealt{Crighton2016, Cooke2017}) or, in some cases, remained chemically pristine (e.g. \citealt{Fumagalli2012, Robert2019}). In this work, we focus on the highest column density systems, $N({\rm H\,\textsc{i}})> 10^{20.3} $cm$^{-2}$, known as Damped Lyman-$\alpha$ systems (DLAs). At these high column densities, the gas is self-shielding; hydrogen is predominantly neutral, while the other elements usually reside in a single, dominant ionisation state. Spectral absorption features associated with the dominant ionic species can therefore be used to determine the relative abundances of elements without the need for ionisation corrections. The oxygen abundance of these systems can be determined reliably because charge-transfer reactions ensure that oxygen closely follows that of hydrogen \citep{FieldSteigman1971}, and we expect dust depletion to be minimal for oxygen (e.g. \citealt{SpitzerJenkins1975}), particularly in the lowest metallicity DLAs\footnote{In addition, provided that an optically thin O~{\sc i} absorption line is available, the determination of the O\,\textsc{i} column density, and hence the oxygen abundance, does not depend on the geometry or kinematics of the gas cloud.}  \citep{Pettini1997,Akerman2005, Vladilo2011, Rafelski2014}. Since oxygen is predominantly sourced from the SNe of massive stars, it can be considered an informative tracer of chemical enrichment \citep{Henry2000}. Throughout this work, we therefore characterise the metallicity of DLAs using their oxygen abundance. \\
  \indent   The most metal-poor DLAs are typically studied at $z\sim3$, when the age of the Universe is $\sim2$ Gyr, therefore, there is a possibility that some of these gas clouds were enriched by subsequent generations of Population II stars. Furthermore, even if all of the metals in near-pristine DLAs come from metal-free stars, it is currently unclear if these metals were produced by stars in the same halo; the minihalos in which the first stars formed are not thought to have evolved into the first galaxies \citep{BrommYoshida2011}. The energetic SNe of the first stars are known to have disrupted the gas within these minihalos --- likely to the point where substantial retention, and subsequent star formation, is implausible \citep{Bromm2003, Greif2007, Greif2010}. Therefore, if the chemical signature of metal-free star formation is detected in near-pristine DLAs, it may have migrated from its initial birthplace, through the intergalactic medium, and into the halos which now host the most metal-poor DLAs. Consequently, the metals in near-pristine gas clouds may represent the combined chemical imprint from multiple minihalos. \\ 
\indent To explore this possibility, and to infer the physical properties of the first stars from the chemistry of EMP DLAs, we require nucleosynthesis simulations that follow the complete chemical evolution of a metal-free star from its initial phases through to the explosive burning phase of its eventual SN explosion. There are several independent groups that have refined this detailed calculation over the years \citep{WoosleyWeaver1995,ChieffiLimongi2004,Tominaga2006,HegerWoosley2010,LimongiChieffi2012}. The relative abundances of metals expelled by the first stars depend on various stellar properties. Parameters commonly considered in the SN calculations include the initial progenitor star mass, the explosion energy, and the mixing between stellar layers. The calculations by \citeauthor{WoosleyWeaver1995} \citepalias[1995; hereafter][]{WoosleyWeaver1995}, \citeauthor{HegerWoosley2010} \citepalias[2010; hereafter][]{HegerWoosley2010}, and \citeauthor{LimongiChieffi2012} \citepalias[2012; hereafter][]{LimongiChieffi2012} all indicate that the ratio of the yields of carbon and oxygen expelled from the SNe of metal-free stars decreases almost monotonically with an increasing progenitor mass. \citetalias{HegerWoosley2010} also find that the ratio of silicon to oxygen, for a given progenitor mass, is sensitive to the explosion energy of the progenitor star.\\
\indent In this paper, we present a novel stochastic enrichment model to investigate the properties on an enriching population of metal-free stars using the relationships found in the \citetalias{HegerWoosley2010} yield set. Our stochastic enrichment model considers the mass distribution of an enriching population as well as the typical SN explosion energy. We employ this model to investigate the enrichment history of the 11 most metal-poor DLAs currently known beyond a redshift of $z=2.6$. 
This analysis complements and extends recent work that approaches this same problem using EMP stars (e.g. \citealt{Ji2015, Fraser2017, Ishigaki2018}). We start by describing our model in Section~\ref{sec:methods}. We summarise the data that are used in our analysis in  Section~\ref{sec:data} and discuss the results of this analysis in Section~\ref{sec:fiducial}. In Section~\ref{sec:disc}, we discuss the possibility of alternative sources of enrichment, the stability of our model, and infer some of the physical properties of the most metal-poor DLAs. We list our main conclusions and discuss the future applications of our model in Section~\ref{sec:conclusions}.

\section{Stochastic Enrichment Model}
\label{sec:methods}
 In this section, we describe our stochastic chemical enrichment model of Population III enriched systems.
Throughout this work we use the definition:
\begin{equation}
\label{eqn:X/Y}
  [{\rm X / Y}] =  \log_{10}\big( N_{{\rm X}}/N_{{\rm Y}}\big) - \log_{10} \big(N_{{\rm X}}/N_{{\rm Y}}\big)_{\odot} \;
\end{equation}
which represents the number abundance ratio of elements X and Y, relative to the solar value. We focus our attention on the [C/O], [Si/O], and [Fe/O] ratios, as these elements are most commonly detected in near-pristine gas. We use the solar ratios as recommended by \cite{Asplund}.
The solar values associated with these elements are\footnote{$\log_{10}\epsilon_{\rm X} = \log_{10}\big(N_{\rm X}/N_{\rm H}\big) + 12$.}: $\log_{10} \epsilon_{\;\rm C_{\odot}} = 8.43$, $\log_{10} \epsilon_{\; \rm O_{\odot}} = 8.69$,  $\log_{10} \epsilon_{ \;\rm Si_{\odot}} = 7.51$, and  $\log_{10} \epsilon_{\; \rm Fe_{\odot}} = 7.47$. \\ 
\indent Relative element abundance ratios can be determined to a precision of $\sim 0.01~{\rm dex}$, provided that the data are collected with a high spectral resolution ($R\gtrsim40,000$) echelle spectrograph and are recorded at signal-to-noise ratio (S/N~$\simeq 15$ per pixel). This high precision allows us to infer the properties of the stars that were responsible for the chemical enrichment of near-pristine gas (e.g. the stellar mass distribution) and the details of the SN explosion that ended the progenitor stars' lives (e.g. kinetic energy, stellar mixing). \\

\subsection{Mass Distribution Model and Likelihood Function}

\label{sec:likelihood}
We model the mass distribution of metal-free stars as a power-law of the form $\xi(M)=k\,M^{-\alpha}$, where $\alpha$ is the power-law slope ($\alpha=2.35$ for a bottom-heavy  Salpeter IMF\footnote{i.e. the first local measurement of the stellar IMF \citep{Salpeter1955}. See \cite{Chabrier2003} for an alternative functional form.}), and $k$ is a multiplicative constant that is set by defining the number of `enriching stars', $N_{\star}$, that form between a minimum mass $M_{\rm min}$ and maximum mass $M_{\rm max}$, given by:
\begin{equation}
\label{eqn:N}
\centering
    N_{\star} = \int_{M_{\rm min}}^{M_{\rm max}} kM^{-\alpha} {\rm d}M  \; .
\end{equation}
In this work, $N_{\star}$ therefore represents the number of stars in this mass range that have contributed to the enrichment of a system, i.e. the `enriching stars'. Note that, in a given metal-poor DLA, the enriching stars may have formed in separate minihalos which later merged or had their chemical products mixed. In this sense, the chemistry of metal-poor DLAs may represent a relatively `well-sampled' IMF of the first stars. In addition to the mass distribution, we also consider the typical SN explosion energy of the enriching stars $E_{\rm exp}$, which is a measure of the kinetic energy of the SN ejecta at infinity. \\
\indent Using a sample of the most metal-poor DLAs, and their constituent abundance ratios, we can investigate the likelihood of a given enrichment model by calculating the probability of the observed abundance ratios, $R_{o}$, given the abundance ratios expected from that enrichment model, $R_{m}$:
\begin{equation}
\label{eqn:like}
        \mathcal{L} = \prod_{n} p_{n} ( R_{o} | R_{m} ) \; ,
    \end{equation}
where $n$ refers to the $n^{\rm th} $ metal-poor DLA in our sample.
The probability of an observed abundance ratio (e.g. [C/O]) is given by 
\begin{equation}
\label{eqn:prob}
    p_{n} \big( R_{o} | R_{m} \big)  = \int  p \big( R_{o} | R_{i} \big)  p \big( R_{i} | R_{m} \big) {\rm d} R_{i}  \; .
\end{equation}
The first term of this integral describes the probability of a given observation being equal to the intrinsic (i.e. true) abundance ratio of the system, $R_{i}$. This distribution is modelled by a Gaussian, where the spread is given by the observational error on the chemical abundance ratio. The second term of the integral in Equation \ref{eqn:prob} describes the probability of obtaining the intrinsic abundance ratio given the IMF defined in Equation \ref{eqn:N} combined with the nucleosynthesis calculations of the ejecta from the enriching stars. Our sample of the most metal-poor DLAs have a minimum of two observed abundance ratios --- both [C/O] and [Si/O] (see Section~\ref{sec:data}). Therefore, in this work, the probability of a system's chemical composition is given by the joint probability of these abundance ratios for a given enrichment model. For the systems that also have an [Fe/O] determination, the probability density is extended to include this ratio as well. \\ 
\indent Our model contains five parameters: $N_{\star}$, $\alpha$, $M_{\rm min}$, $M_{\rm max}$, and $E_{\rm exp}$. In the case of a well-sampled IMF, $R_{i}=R_{m}$; however, as the first stars are thought to form in small multiples \citep{Turk2009, Stacy2010}, the number of enriching stars is expected to be small. Thus, the IMF of the first stars is stochastically sampled. Due to the stochasticity of the IMF, we have to construct abundance ratio probability distributions, $p (R_{i} | R_{m})$, for each combination of our fiducial model parameters. The range of model parameters we consider are:
\begin{eqnarray*}
 1 \leq &N_{\star}& \leq 100 \; , \\
 -5 \leq &\alpha &\leq 5 \; , \\
20\leq &M_{\rm max}/{\rm M_{\odot}}& \leq 70 \; , \\
 0.3 \leq &E_{\rm exp}/10^{51}{\rm erg}& \leq 10  \; .
\end{eqnarray*}
In what follows, we assume that stars with masses $>10~{\rm M_{\odot}}$ are physically capable of undergoing core-collapse. Therefore, this parameter is fixed at a value $M_{\rm min}=10~{\rm M_{\odot}}$.
We also consider a maximum mass, $M_{\rm max}$, above which all stars are assumed to collapse directly to a black hole, and do not contribute to the chemical enrichment of their surroundings. We impose a uniform prior of $20 < M_{\rm max}/{\rm M_{\odot}} < 70$ on the maximum mass of the enriching stars --- this upper bound corresponds to the mass limit above which pulsational pair-instability SNe are believed to occur \citep{Woosley2017}. Similarly, we impose a uniform prior on the explosion energy, a choice that is driven by the yield set utilised in this analysis. We describe these nucleosynthesis yields in more detail in the following section. The explored range of $E_{\rm exp}$ covers all feasible explosion energies given our current understanding of core-collapse SNe.

\subsection{Ejecta of Metal-Free Stars}
\label{sec:yields}
\begin{figure}
\centering
\includegraphics[width = 0.9\columnwidth]{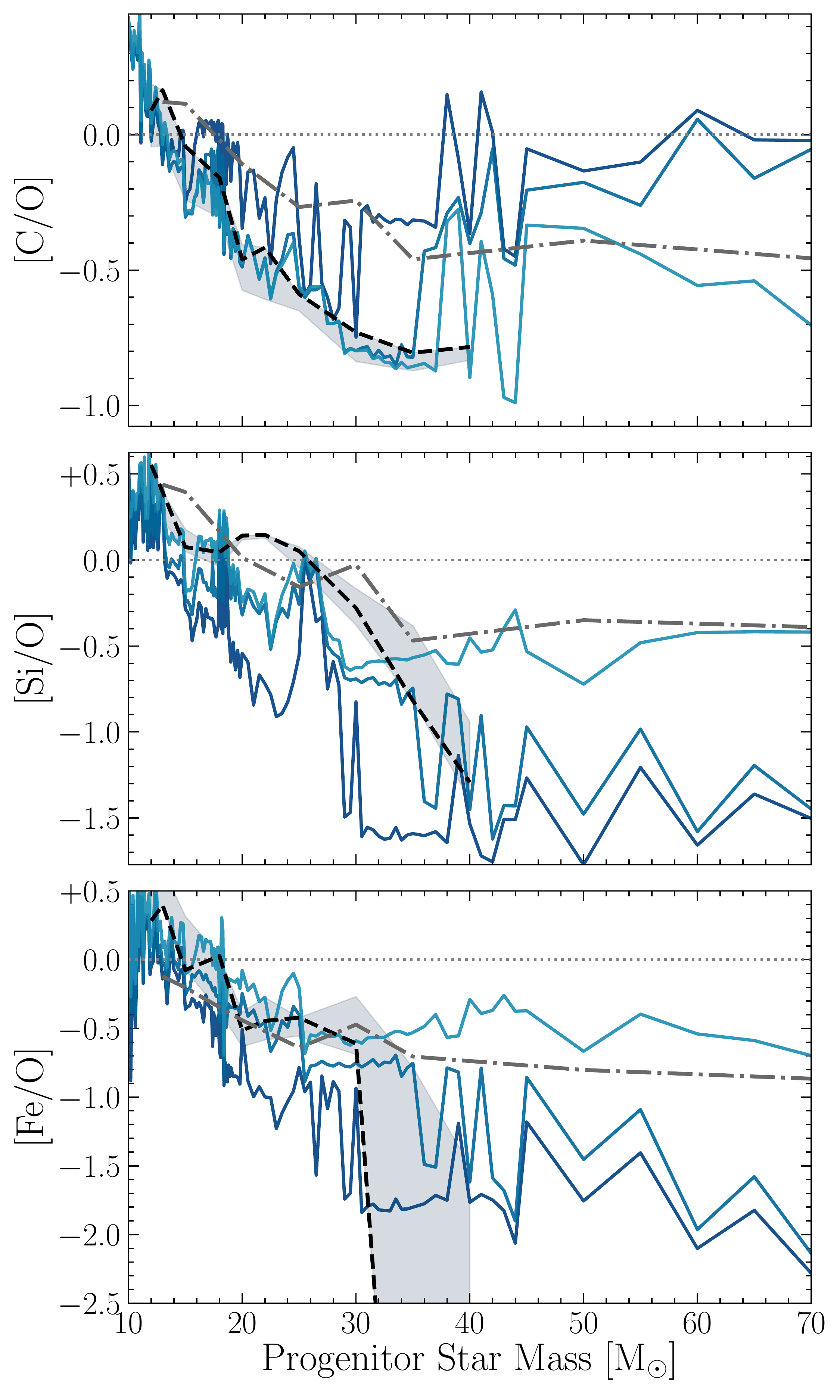}
\caption{Relationship between the ejected [C/O], [Si/O], and [Fe/O] abundance ratios as a function of the stellar progenitor mass for a range of explosion energies. The dark blue line corresponds to a $\rm 1.2~B$ explosion while the progressively lighter lines correspond to a $\rm 1.8~B$ and $\rm 5~B$ explosion respectively (note ${\rm1~B = 10^{51} erg}$). Yields are taken from \protect\citetalias{HegerWoosley2010} and are shown relative to the solar values recommended by \protect\citealt{Asplund}. The solar abundances are marked by the horizontal dotted grey line. The dashed black curves show the abundance ratios expected from a progenitor with a metallicity $10^{-4}$ that of the Sun ($Z_{\odot}$); these yields are taken from \protect\citetalias{WoosleyWeaver1995} for a typical explosion energy of $\rm 1.2~B$. Also shown, via the dot-dashed grey lines, are the yields of massive metal-free stars as calculated by \protect\citetalias{LimongiChieffi2012}. The explosion energy associated with these progenitors is $\rm \sim 1~B$, however the precise value varies with progenitor mass (as for the \protect\citetalias{WoosleyWeaver1995} yields). The grey-shaded region encompasses the yields expected from all stars in a metallicity range of $10^{-4}< Z/{\rm Z}_{\odot}< 1$, based on the yields computed by \protect\citetalias{WoosleyWeaver1995}. }
\label{fig:CO_v_explos_eng}
\end{figure}
Our analysis relies on simulations of the evolution and eventual SN explosions of massive metal-free stars.
In our work, we adopt the \citetalias{HegerWoosley2010} yields as our fiducial model and utilise the yields of \citetalias{WoosleyWeaver1995} and \citetalias{LimongiChieffi2012} as points of comparison. In \citetalias{HegerWoosley2010}, the nucleosynthetic yields of elements expelled from the SNe of massive metal-free stars are calculated as a function of the progenitor star mass, explosion energy, and the degree of mixing between the stellar layers. \\
\indent The main impediment to the rigour of these SNe yield calculations is the uncertainty surrounding the ultimate explosion of a massive star (e.g. \citealt{Melson2015}). To overcome this, the simulations are performed in one dimension and the explosion is parameterised by a mixing prescription combined with a piston (i.e. a time-dependent momentum deposition that is characterised by a final kinetic energy of the ejecta at infinity, $E_{\rm exp}$). In \citetalias{HegerWoosley2010}, the width of the mixing region is defined as a fraction of the He core size. Their simulations consider 14 mixing widths. However, they recommend adopting a width that is 10 per cent of the He core size, as this provides the best fit to observations of the light curve of SN 1987A. These model yields have been found to provide good fits to the abundance patterns of EMP stars, specifically those from \citet{Cayrel2004}. However, we note that to properly account for mixing driven by Rayleigh-Taylor instabilities and rotation it is necessary to perform these simulations in two or three spatial dimensions (e.g. \citealt{Joggerst2010-ROT, Joggerst2010-R-T,Vartanyan2018}). Further simplifications arise from performing these simulations in isolation, for non-rotating stellar models with negligible magnetic fields and no mass loss. The incorporation of rotation has been shown to induce additional mixing between stellar layers and lead to modest mass loss \citep{Ekstrom2008}. Work by \cite{Yoon2012} has suggested that this mass loss increases in the presence of magnetic torques. \\
\indent  The parameter space explored by \citetalias{HegerWoosley2010} spans masses from $(10-100)~{\rm M_{\odot}} $ and explosion energies from $(0.3-10)\times10^{51}$ erg. This space is evaluated across 120 masses and 10 explosion energies. The average mass spacing between successive yield calculations is $<1$M$_{\odot}$ (and in some cases, as low as $0.1\,{\rm M}_{\odot}$). For comparison, the average mass spacing in \citetalias{LimongiChieffi2012} is $>4$M$_{\odot}$. As can be seen from Figure~\ref{fig:CO_v_explos_eng}, the ejected yields fluctuate rapidly across a small range of progenitor star masses. The \citetalias{HegerWoosley2010} calculations are the only yield calculations with a mass spacing small enough to account for this behaviour, which is thought to arise due to the non-linear interactions between the burning shells within a star \citep{Muller2016, Sukhbold2018}. Utilising the \citetalias{HegerWoosley2010} yields enables us to investigate the properties of our enriching stars with a finer mass resolution than would be afforded by other yield models. \\
\indent As can be seen from Figure~\ref{fig:CO_v_explos_eng}, the [C/O] abundance ratio evolves almost monotonically with progenitor mass for stars that explode with an energy $\rm \gtrsim1.8~B$ and are $<40~{\rm M_{\odot}}$. The shells in which carbon and oxygen form are relatively close to the surface of a star; for explosions above $\sim1.8\rm ~B$, these outer layers are mostly ejected. However, elements closer to the iron peak, like silicon and iron, are more dependent on the energy of the explosion, and are more likely to fall back onto the newly formed central compact object. Therefore, the combined analysis of the [C/O] and [Si/O] ratios of a system enriched by one SN would place constraints on the mass and explosion energy of the enriching star. Section~\ref{sec:PDFs} describes how we extend this to systems that have been enriched by a small number of stars, as opposed to just a single star. \\
\indent As a point of comparison, the grey shaded regions in Figure~\ref{fig:CO_v_explos_eng} indicate the yields of massive Population II and Population I stars calculated by \citetalias{WoosleyWeaver1995}. This comparison suggests that the relative yields of the most abundant elements are almost indistinguishable between metal-free and metal-enriched massive stars. Given the similarity of the yields of these elements across different stellar populations, we use the \citetalias{HegerWoosley2010} models owing to their fine mass resolution and the large grid of explosion energies, regardless of whether the most metal-poor DLAs were enriched by Pop. III or Pop. II stars. We consider the potential of enrichment from alternative sources in Section~\ref{sec:disc}.
\begin{figure}
    \centering
    \includegraphics[width = .9\columnwidth]{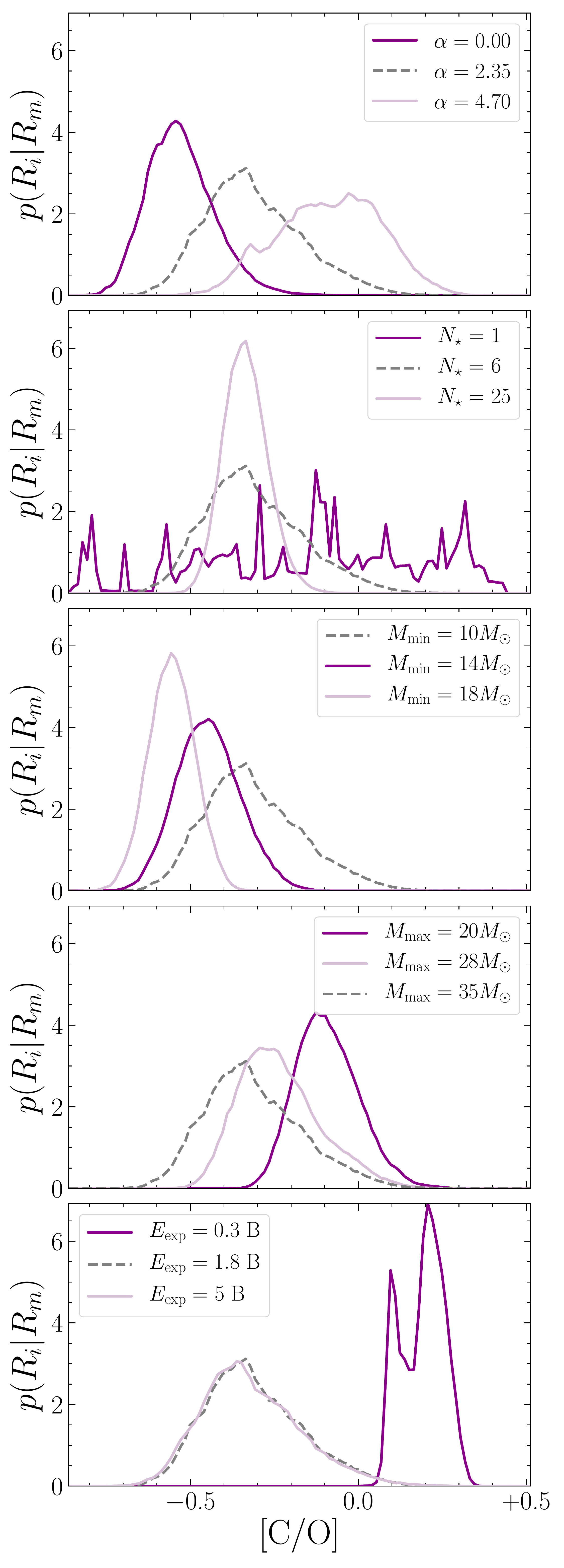}
    \caption{The [C/O] distribution for a range of enrichment models. The successive panels correspond to changing the slope, number of enriching stars, minimum mass, maximum mass, and explosion energy, respectively. Unless otherwise stated in the legend, the model parameters of these distributions are $\alpha = 2.35$, $N_{\star} = 6$, $M_{\rm min} = 10$ M$_{\odot}$, $M_{\rm max} = 35$ M$_{\odot}$, and $E_{\rm exp} = 1.8$ B (displayed as the grey dashed line in all panels as a point of comparison).}
    \label{fig:CO_PDF}
\end{figure}

\begin{table*}

    \caption{Abundance ratios of metal-poor gas clouds with known hydrogen column densities}
    \label{tab:CO}
    \begin{tabular}{lcccccccc }  
        \hline
        QSO & $z_{\rm abs}$ & $\log_{10}$~N(H~{\sc i}) & [Fe/H] & [O/H] & [C/O] & [Si/O]  & [Fe/O]  & References \\
        \hline
J0140--0839  & 3.6966 &     20.75 & $-3.45 \pm  0.24$       &   --2.75 $\pm$ 0.05     &    --0.30 $\pm$ 0.08    &    0.00  $\pm$ 0.09  & --0.70 $\pm$ 0.19  &  1,2    \\
J0311--1722  & 3.7340 &     20.30 & $< -2.01$               &   --2.29 $\pm$ 0.10     &    --0.42 $\pm$ 0.11    &  --0.21  $\pm$ 0.11  & $< +0.28$           &  2    \\
J0903+2628   & 3.0776 &     20.32 & $< -2.81$               &   --3.05 $\pm$ 0.05     &    --0.38 $\pm$ 0.03    &  --0.16  $\pm$ 0.02  & $< +0.24$           &  3    \\
Q0913+072    & 2.6183 &     20.34 & $-2.82 \pm 0.04$        &   --2.40 $\pm$ 0.04     &    --0.36 $\pm$ 0.01    &  --0.15  $\pm$ 0.01  & --0.42 $\pm$ 0.04  &  4,5  \\
J0953--0504  & 4.2029 &     20.55 & $-2.95 \pm 0.21$        &   --2.55 $\pm$ 0.10     &    --0.50 $\pm$ 0.03    &  --0.16  $\pm$ 0.03  & --0.40 $\pm$ 0.22  &  6    \\
J1001+0343   & 3.0784 &     20.21 & $-3.18 \pm 0.15$        &   --2.65 $\pm$ 0.05     &    --0.41 $\pm$ 0.03    &  --0.21  $\pm$ 0.02  & --0.53 $\pm$ 0.15  &  2    \\
J1016+4040   & 2.8163 &     19.90 & \ldots                  &   --2.46 $\pm$ 0.11     &    --0.21 $\pm$ 0.05    &  --0.05  $\pm$ 0.06  & \ldots             &  5    \\
Q1202+3235   & 4.9770 &     19.83 & $-2.44  \pm 0.16$       &   --2.02 $\pm$ 0.13     &    --0.33 $\pm$ 0.11    &  --0.43  $\pm$ 0.09  & --0.42 $\pm$ 0.18  &  7    \\
J1337+3153   & 3.1677 &     20.41 & $-2.74  \pm  0.30$      &   --2.67 $\pm$ 0.17     &    --0.19 $\pm$ 0.11    &  --0.01  $\pm$ 0.10  & --0.07 $\pm$ 0.31  &  8    \\
J1358+6522   & 3.0673 &     20.50 & $-2.88  \pm 0.08$       &   --2.34 $\pm$ 0.08     &    --0.27 $\pm$ 0.06    &  --0.23  $\pm$ 0.03  & --0.54 $\pm$ 0.08  &  4,9  \\
J2155+1358   & 4.2124 &     19.61 & $-2.15  \pm 0.25$       &   --1.80 $\pm$ 0.11     &    --0.29 $\pm$ 0.08    &  --0.07  $\pm$ 0.06  & --0.35 $\pm$ 0.26  &  10   \\
\hline
\end{tabular} \\
 1: \cite{Ellison2010}, 2: \cite{Cooke2011}, 3: \cite{Cooke2017},  4: \cite{Cooke2014}, 5: \cite{Pettini2008}, 6: \cite{Dutta2014}, 7: \cite{Morrison2016}, 8: \cite{Srianand2010}, 9: \cite{CookePettiniMurphy2012}, 10: \cite{Dessauges-Zavadsky2003}
\end{table*}

\subsection{Model Abundance Ratios}
\label{sec:PDFs}
The first stars likely formed in small multiples, which necessarily means that their underlying mass distribution is stochastically sampled. To account for this, we construct abundance ratio probability distributions using Monte Carlo simulations. For a given IMF model, we stochastically sample the distribution and use the resulting progenitor star masses to calculate the total yield of C, O, Si, and Fe, based on the \citetalias{HegerWoosley2010} simulations. For the case of [C/O], the total yield of carbon and oxygen supplied by all of the stars is used to determine the resulting number abundance ratio:
\begin{equation}
\label{eqn:NC_NO}
N_{\rm C} / N_{\rm O} = m_{\rm O}/m_{\rm C}\frac{\sum\limits^{N_{\star}}_{i=1}M_{\mathrm{C},i}}{\sum\limits^{N_{\star}}_{i=1}M_{\mathrm{O}_,i}}
\end{equation}
where $m_{\rm C}$ and $m_{\rm O}$ are the masses of a single carbon and oxygen atom, respectively; $M_{\mathrm{ C}, i}$ and $M_{\mathrm{O}, i}$ are the masses of these elements that are expelled from the SN of star $i$ within the multiple. From this we obtain a stochastically sampled [C/O] ratio. This is repeated $10^{3}$ times to construct the probability density function, $p (R_{i} | R_{m})$ in Equation~\ref{eqn:prob}, of [C/O] for a given mass distribution model and explosion energy. In actuality this sampling procedure is performed for [C/O], [Si/O], and [Fe/O] simultaneously and we consider the 3D joint probability density function of all of the ratios. In Figure~\ref{fig:CO_PDF} we have marginalised over both [Si/O] and [Fe/O] to illustrate the sensitivity of each model parameter to the resulting [C/O] distribution. The successive panels correspond to changing the slope, number of enriching stars, minimum mass, maximum mass, and explosion energy respectively. The example model parameters used in Figure~\ref{fig:CO_PDF} (grey-dashed curves) are: $\alpha = 2.35$, $N_{\star} = 6$, $M_{\rm min} = 10~{\rm M_{\odot}}$, $M_{\rm max} = 35~{\rm M_{\odot}}$, and $E_{\rm exp}=1.8~{\rm B}$. Note that when we compare the observed abundance ratios of a sample of systems to those from the adopted \citetalias{HegerWoosley2010} yields, we are assuming that all of the SNe that enriched these systems are well-modelled by the same explosion energy. It is likely that SNe with a range of explosion energies contributed to the enrichment of metal-poor DLAs. Due to computational limitations, we cannot treat the explosion energies of individual stars stars independently; our chosen prescription should therefore be considered to represent the `typical' $E_{\rm exp}$ of the enriching stars. In the future we may consider a mass dependent explosion energy. However, the present generation of explosive nucleosynthesis models are not quite at the point whereby the kinetic energy released by the SN explosion is known as a function of the progenitor mass. Indeed, the expected functional form may not be parametric at all; recent calculations suggest that there are `islands of explodability' for massive stars (e.g. \citealt{Sukhbold2018}). Furthermore, the latest models of core-collapse SNe by \cite{Muller2019} indicate that $\sim10~{\rm M}_{\odot}$ stars tend to yield somewhat low kinetic energy ($\sim0.3~{\rm B}$). Given the uncertainty surrounding the appropriate parameterisation, we favour our chosen prescription due to its simplicity and reserve the consideration of alternatives for future investigations. We can nevertheless consider how our assumption might impact our inferred parameter values. For a given value of $N_{\star}$, our model assumes that all stars explode with the same final kinetic energy at infinity. If we were to allow every star to explode with a different energy, this likely produces a greater diversity of the element abundances ratios, thus broadening the $p (R_{i} | R_{m})$ distribution. As the second panel of Figure~\ref{fig:CO_PDF} highlights, reducing the number of stars that have enriched a system also broadens the distribution of allowed abundances. Consequently, we may infer a lower $N_{\star}$ to account for the spread of a given abundance observed within our sample. \\
\indent One of the underlying assumptions of Equation~\ref{eqn:NC_NO} is that the metals ejected by the first stars were uniformly mixed. Considering the time between the first episodes of enrichment and when the metal-poor DLAs in our sample have been observed, it is likely that the enriched gas within these systems has had sufficient time to become well-mixed (see e.g. \citealt{Webster2015}). In any case, when we measure the relative element abundances of a gas cloud, we are taking the average across the entire sightline. Therefore, the measured abundance ratio of a given gas cloud should be representative of the number ratio in Equation~\ref{eqn:NC_NO} even if it contains pockets of unmixed SNe ejecta.

\subsection{Likelihood Sampling Technique}
The likelihood function (Equation~\ref{eqn:like}) is sampled using a Markov Chain Monte Carlo (MCMC) procedure. We utilise the \textsc{emcee} software package \citep{EMCEE} for this purpose. We draw $8.4\times10^{5}$ samples across 400 walkers each taking 2100 steps. We adopt a conservative burn-in that is half the length of the original chains. We consider the chains converged once doubling the number of steps taken by each walker has no impact on the resulting parameterdistributions.
We also repeat the analysis using a different seed to generate the initial randomised positions of the walkers. As our results do not change, this provides an additional test of convergence. We display the results of our MCMC analysis using the \textsc{corner} software package \citep{CORNER}. To check that our results are not driven by a single observational data point within our sample we perform a bootstrap analysis to gauge the sampling error associated with our maximum likelihood estimates. 
\begin{figure*}
    \centering
    \hspace*{-1.5cm}
    \includegraphics[width = 1.2\textwidth]{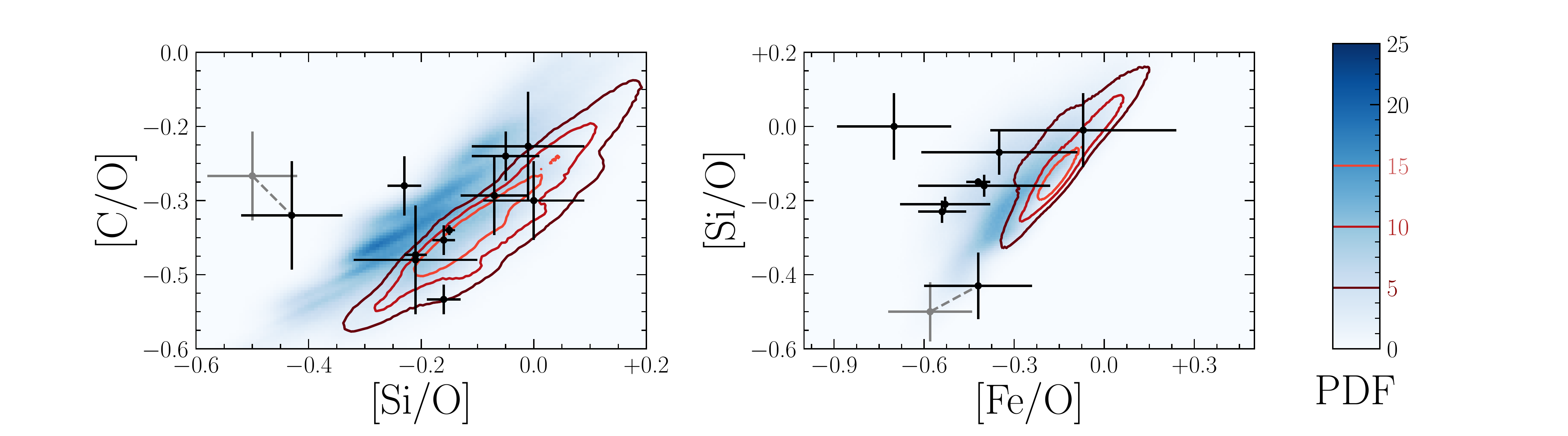}
    \caption{The [C/O], [Si/O], and [Fe/O] abundance ratios of the high redshift ($z\geq2.6$) systems used in our analysis (black symbols with error bars) overplotted on the joint probability distributions (blue shaded distributions) of [C/O] and [Si/O] (left) and [Si/O] and [Fe/O] (right) of an example model where $\alpha = 2.35$, $N_{\star} = 6$, $M_{\rm min} = 10$ M$_{\odot}$, $M_{\rm max} = 35$ M$_{\odot}$, and $E_{\rm exp} = 1.8$ B (i.e. the same example model shown by the grey dashed curves in Figure~\ref{fig:CO_PDF}). The red contours show the same joint probability distributions for the case of a 5~B  explosion. The colours of the contours correspond to the probability region they encompass (as indicated on the colourbar). The grey data points highlight the abundance ratios of the sub-DLA along the line-of-sight towards Q1202+3235 adopted by \protect\citet{Morrison2016}. The grey dashed lines connect these data to that adopted in this work (black symbols; see text).}
    \label{fig:CO_SiO_PDF_data}
\end{figure*}

\section{Data}
\label{sec:data}

\indent Our sample consists of the abundance ratios of the most metal-poor DLAs currently known. Specifically, that of [C/O], [Si/O], and when available,  [Fe/O]. These abundances have been determined from high resolution spectra taken with either the ESO Ultraviolet and Visual Echelle Spectrograph (UVES; \citealt{Dekker2000}) or the Keck High Resolution Echelle Spectrometer  (HIRES; \citealt{Vogt1994}). 
In Table~\ref{tab:CO}, we list the chemical abundances of these systems.\\
\indent We focus on DLAs with a redshift of $z\geq2.6$ to minimise the potential for enrichment from later generations of star formation (Welsh et al, in prep). Other possible sources of enrichment will be discussed in Section~\ref{sec:disc}. Figure~\ref{fig:CO_SiO_PDF_data} shows the joint [C/O], [Si/O], and [Fe/O] abundance ratios of the systems in our sample. These data are overplotted on the joint probability distribution of [C/O] versus [Si/O] (left) and [Si/O] versus [Fe/O] (right) given the same example model shown in Figure~\ref{fig:CO_PDF}. To offer a point of comparison, we also overplot the abundance ratio distributions for an explosion energy of 5 B (red contours), with all other model parameters unchanged. This highlights the sensitivity of [Si/O] and [Fe/O] to the explosion energy. In this figure, we display two different abundance determinations of the sub-DLA  at $z_{\rm abs}=4.9770$ along the line-of-sight to the quasar Q1202+3235. The authors of the discovery paper \citep{Morrison2016} model the absorption system with multiple velocity components. Some of these velocity components show  C\,{\sc ii} and Si\,{\sc ii} absorption features without corresponding O\,{\sc i} absorption, indicating the presence of ionised gas. \cite{Morrison2016} measure the total element column densities of the system by summing over all of the velocity components and performing ionisation corrections. Instead, we prefer to solely consider the uncorrected column densities of the {\it primary} velocity component, which shows corresponding absorption from O\,{\sc i},  C\,{\sc ii}, and Si\,{\sc ii} (i.e. the absorption component  at $z_{\rm abs}=4.977004$, which is the absorption predominantly arising from neutral gas). In each panel of Figure~\ref{fig:CO_SiO_PDF_data}, the chemical abundance reported by \cite{Morrison2016} is shown by a grey symbol and is connected to our determination (black symbol) by a grey dashed line. Our determination results in a lower [C/O] and [Fe/O] ratio as well as a higher [Si/O] ratio. \\

\section{fiducial model analysis}
\label{sec:fiducial}
\begin{figure*}
\centering
    \includegraphics[width =\textwidth]{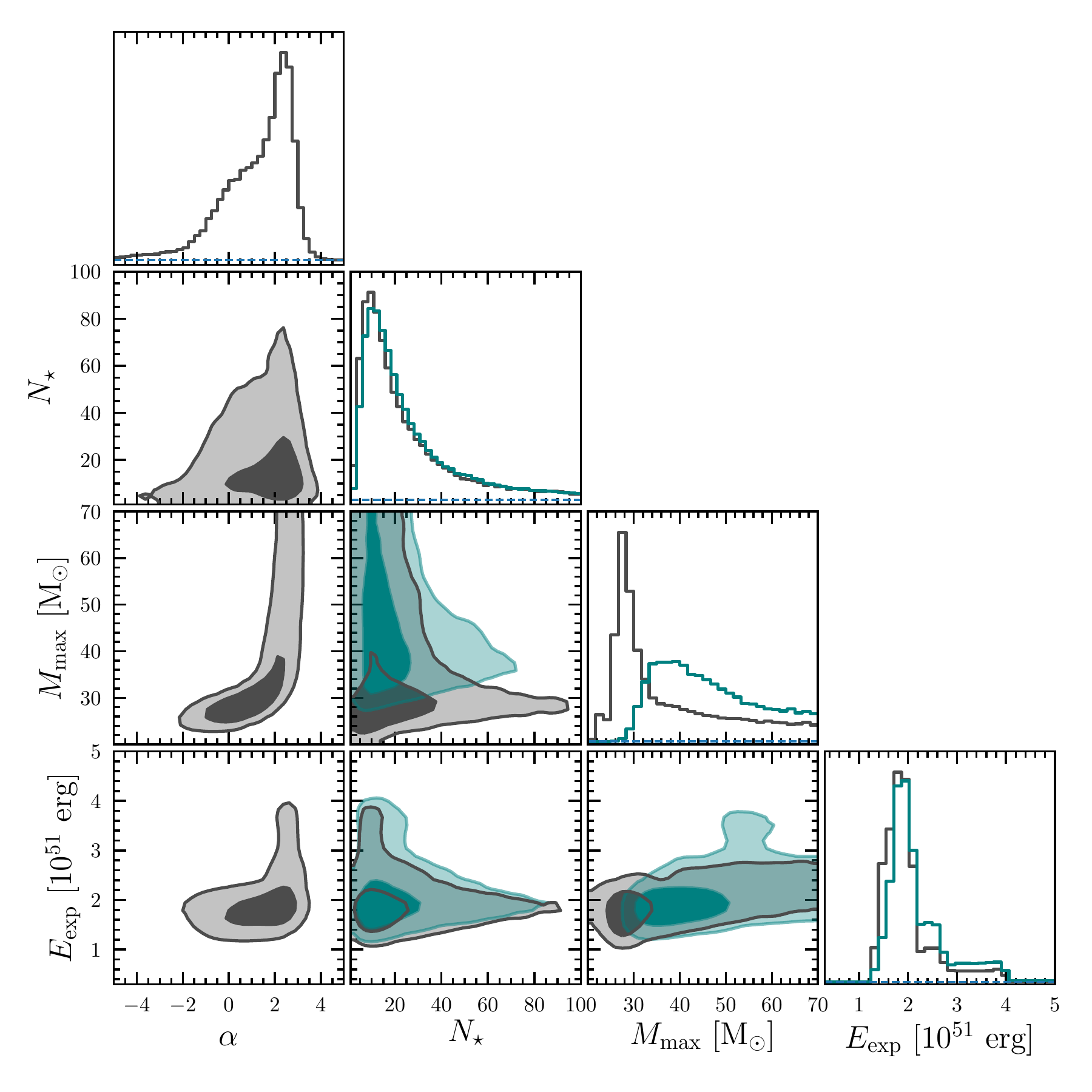}
    \caption{The marginalised maximum likelihood distributions of our fiducial model parameters (main diagonal), and their associated 2D projections, given the high redshift, metal-poor DLAs listed in Table \ref{tab:CO}. The dark and light contours show the 68 per cent and 95 per cent confidence regions of these projections respectively. The horizontal blue dashed lines mark where the individual parameter likelihood distributions fall to zero. The grey distributions correspond to the analysis of the full parameter space, described in Section~\ref{sec:methods}. The green distributions are the result of imposing a Salpeter slope for the IMF (i.e. $\alpha=2.35$).}
    \label{fig:corner_fid}
\end{figure*}

\indent Our fiducial model assumes that stars with masses above $10~{\rm M_{\odot}}$ can undergo core-collapse. We therefore impose a hard prior on the lower mass bound of our model IMF. The remaining model parameters are free to vary within limited bounds, as described in Section~\ref{sec:methods}. In Figure~\ref{fig:corner_fid}, we show the posterior distributions (black histogram; diagonal panels) and 2D projections (grey contours) of the model parameters, based on the 11 most metal-poor DLAs at redshift $z\geq2.6$.
In the following subsections we discuss each parameter distribution individually. Throughout, the quoted errors on our maximum likelihood estimates are found using a bootstrap analysis of these data. The errors indicate the stability of our maximum likelihood estimates by measuring the variability of this statistic across multiple data realisations. Specifically, they are the 68 per cent confidence regions around the median maximum likelihood estimates across all bootstraps.
\subsection{Slope}

The maximum likelihood estimate of the slope parameter is $\hat{\alpha}=2.5 \pm 0.2$. Within the bootstrapped errors, this estimate encompasses a Salpeter distribution. Our result is therefore consistent with empirical determinations of the power-law slope of the IMF at $M\gtrsim1~{\rm M_{\odot}}$. However, there is a broad tail towards a flatter, and even top heavy, slope. Given the broad range of $\alpha$ values recovered by our fiducial model, we have recalculated the results after imposing a strong Salpeter-like prior on the slope parameter, $\alpha=2.35$. The result of this analysis is overplotted in Figure~\ref{fig:corner_fid} (green contours). The distributions of $N_{\star}$ and $E_{\rm exp}$ are virtually unchanged under the assumption of a Salpeter IMF, while the $M_{\rm max}$ distribution is broadened and shifted towards a higher mass limit. This suggests that the enrichment of the systems in our sample can be well-described by stars drawn from a Salpeter-like IMF. \\
\indent A Salpeter-like IMF could indicate that the chemical signature of the DLAs in our sample are dominated by the contribution from second generation stars. However, further work is needed to distinguish the signature of Population II versus Population III enrichment. To isolate the chemical signature of the first stars, we should restrict our analysis to the {\it most} metal-deficient DLAs, ideally, those with [O/H] $<-3$. Currently, there are not enough systems known within this regime to implement such an analysis --- only one system, J0903+2628, has been found with an oxygen abundance [O/H] $<-3$ \citep{Cooke2017}. \\

\subsection{Enriching Stars}
As can be seen in the second panel of Figure~\ref{fig:CO_PDF}, the \emph{intrinsic} spread of the relative element abundance ratios is sensitive to the sampling of the IMF. Specifically, the distribution of [C/O] becomes more centrally concentrated as more stars enrich each system (i.e. in the limit of a well-sampled IMF, all DLAs would exhibit an almost identical [C/O] ratio). Thus, if the scatter between the data points is larger than the quoted errors, then we can use the scatter to probe the sampling of the IMF. For our fiducial model, the maximum likelihood estimate of the number of enriching stars is $\hat{N}_{\star}=10 \pm 4$.  The $95^{\rm th}$ percentile of this distribution suggests $N_{\star} \lesssim 72$. These statistics are unchanged under the assumption of a Salpeter IMF. From this, we conclude that a typical DLA in our sample has been enriched by a small number of  massive stars.

\subsection{Maximum Mass}
The maximum likelihood estimate of the upper mass limit of enriching stars is $\hat{M}_{\rm max}=(28 \pm 1) \,{\rm M_{\odot}}$. 
The interquartile range of this distribution spans $(28-45)\,{\rm M_{\odot}}$. 
As can be seen in Figure~\ref{fig:corner_fid}, the posterior distribution on $M_{\rm max}$ has a broad tail towards high progenitor masses. This should be expected, since the data are consistent with a bottom-heavy IMF. This means that stars preferentially form with lower masses, and higher mass stars are not well-sampled. As the most massive stars have a low occurrence rate, it is difficult to discern the maximum cutoff mass, above which stars do not contribute to the enrichment of metal-poor DLAs. In the case of a Salpeter IMF, the maximum likelihood estimate of $M_{\rm max}$ shifts to a larger value ($\approx 40~{\rm M_{\odot}}$) and the overall distribution becomes broader. \\
\indent Regardless of whether we impose a prior on the slope parameter, the maximum likelihood estimate of the upper mass limit is $<40~{\rm M_{\odot}}$. This limit was also reported by \cite{Ishigaki2018} who investigated the chemical enrichment of metal-poor halo stars. Our results tentatively support the work of \cite{Sukhbold2016} (see also, \citealt{Burrows2019}). These authors found that, when an explosion model is powered by neutrinos, only a fraction of the stars above $20~{\rm M_{\odot}}$ have sufficient energy to successfully launch a SN explosion. The remaining stars are presumed to collapse directly to black holes. Recent work by \cite{Sukhbold2019_pre} suggests that the apparent mass dependence of a progenitor's `explodability' may be the consequence of a transition in the dominant carbon burning regime that occurs within the presupernova cores of progenitors at $\sim20~{\rm M_{\odot}}$. This scenario is supported observationally by \cite{Adams2016}, who have identified a potential failed SN in the form of a star that disappeared from multi-epoch LBT imaging;
a technique envisioned by \cite{Kochanek2008} and later implemented by \cite{Gerke2015} and \cite{Reynolds2015}. We note that this result is also consistent with \citet{Heger2003}, which reports the direct collapse of metal-free stars above $40~{\rm M_{\odot}}$.

\subsection{Explosion Energy}
The maximum likelihood estimate of the typical explosion energy is $\hat{E}_{\rm exp}=1.8^{+0.3}_{-0.2}\times10^{51}~{\rm erg}$. Under the assumption of a Salpeter IMF, $\hat{E}_{\rm exp} \approx 2\times10^{51}~{\rm erg}$, which is consistent with the results of our fiducial model within the bootstrapped error bounds. The distribution of this parameter is the most well-defined, with an interquartile range spanning $(1.7-2.1)\times 10^{51}~{\rm erg}$. \\
\indent Our inferred enrichment model indicates that it is the lowest mass progenitors that are responsible for the enrichment of the DLAs in our sample. For these stars, simulations predict $<10^{51}~{\rm erg}$ explosions (e.g. \citealt{Muller2019}). The high [Fe/O] yields associated with the high energy explosions of the lowest mass progenitors from \citetalias{HegerWoosley2010} may therefore be unrepresentative of a realistic scenario. It is these high [Fe/O] yields of the lowest mass progenitors that drive our analysis to disfavour models with high typical explosion energies (see the red contours in Figure~\ref{fig:CO_SiO_PDF_data} for an example of how an increase in the explosion energy impacts the expected range of observed abundances). We find it encouraging that our analysis shows no evidence for the models disfavoured by these simulations. As mentioned in Section~\ref{sec:PDFs}, a potential future avenue of investigation is the consideration of a mass dependent explosion energy model; this may help accommodate the behaviour seen in recent simulations \citep{Muller2019}. \\
\indent In \citetalias{HegerWoosley2010}, the authors found that the abundance patterns of EMP halo stars (i.e. the \citealt{Cayrel2004} sample) are best described by enrichment from SNe, typically with $0.6\lesssim E_{\rm exp}/10^{51}~{\rm erg} \lesssim1.2$. In contrast to this, a similar analysis performed by \cite{GrimmettHeger2018} found that the abundance patterns of EMP halo stars are best described by the yields of $(5-10)\times10^{51}~{\rm erg}$  explosions (i.e hypernovae). This preference towards enrichment by a population of high energy SNe was also reported by \cite{Cooke2017}  and  \cite{Ishigaki2018}. Furthermore, the observed overabundance of [Zn/Fe] in the most metal-poor halo stars \citep{Primas2000, Cayrel2004}, is thought to be due to enrichment by a population of hypernovae \citep{Umeda2002}. Although the explosion energy that we derive in this work is somewhat lower than that found in other studies, our DLA sample probes a somewhat higher metallicity regime, $-3.0\lesssim{\rm [O/H]}\lesssim-2.0$, where metal-poor stars exhibit solar relative abundances of [Zn/Fe]. The metal-poor DLAs in our sample may therefore be displaying the signature of enrichment from massive Population II stars that ended their lives with more moderate energy SN explosions.\\

\section{Discussion}
\label{sec:disc}
In the previous section, we investigated the properties of a metal-free stellar population that can describe the chemical abundance patterns of the metal-poor DLA population. The results of our fiducial model analysis suggest that the DLA abundances are well-described by enriching stars drawn from a Salpeter-like IMF at $M>10~{\rm M_{\odot}}$. These results also suggest that a typical metal-poor DLA has been enriched by $\lesssim72$ massive stars (95 per cent confidence) and that these gas clouds have not been significantly enriched by stars with masses $\gtrsim40~{\rm M_{\odot}}$. The ability to recover a constraint on the IMF slope through the analysis of 11 systems is an encouraging sign that this model is a powerful tool. We find that the potential of this analysis is maximised when we demand that a given enrichment model is able to simultaneously reproduce all of the abundance ratios observed within a system. \\
\indent In this section, we discuss the impact of alternative enrichment sources and the sensitivity of our results to the choice of chemical yields. We also highlight some of the inferences that can be made about metal-poor DLAs given an appropriate enrichment model. 

\subsection{Alternative Enrichment Sources}
\label{sec:contam}
As mentioned in Section~\ref{sec:data}, we restrict our analysis to systems found beyond a redshift of $z=2.6$ to minimise the potential for enrichment from non-Pop III stars (Welsh et al. 2019, in prep.). However, given that second generation stars are expected to have formed before this epoch, we must consider avenues through which metal-enriched stars can wash out the signature of Population III stars in the most metal-poor DLAs. Possible mechanisms include: 
\begin{enumerate}
    \item Mass loss from Asymptotic Giant Branch (AGB) stars,
    \item Type Ia SNe ejecta, and
    \item Population II core-collapse (Type II) SNe ejecta.
\end{enumerate}
We now discuss each of these possible enrichment avenues in turn.
\subsubsection{AGB stars}
Intermediate mass ($1-6~{\rm M_{\odot}}$) Population II stars are capable of producing a significant quantity of carbon during their AGB phase \citep{Karakas2014, Hofner2018}. In what follows, we use the model parameter distributions of our fiducial model (with a prior $\alpha=2.35$; green histograms in Figure~\ref{fig:corner_fid}) to estimate the number of AGB stars that may have contributed to the enrichment of the metal-poor DLAs in our sample.  Using a similar approach to that adopted in Section~\ref{sec:methods}, we then perform Monte Carlo simulations to sample stars within the AGB mass range. The carbon lost by these stars is determined using the AGB yield calculations performed by \cite{Karakas2007} and updated in \cite{Karakas2010}. Comparing the distribution of carbon expected from AGB stars to that expected from massive metal-free stars, we find that AGB stars can match ($\approx~110$ per cent) the carbon yield from massive stars. The yields of all other elements considered in our analysis are negligible. For this estimate, we only consider the contribution from stars with masses $M>2~{\rm M_{\odot}}$ since lower mass stars have lifetimes in excess of 2~Gyr; given that Population II stars likely formed at $z<10$, stars with $M\lesssim2~{\rm M_{\odot}}$ will still be on the main sequence when most of the DLAs in our sample are observed (typically $z\sim 3$). Note, the contribution of carbon from Population II AGB stars would be even less if these stars were born more recently than $z\simeq10$. To estimate how the presence of AGB stars could impact our inferences, we have repeated our analysis under the assumption that half of the carbon in a given system can be attributed to AGB stars. We find a preference towards both higher typical explosion energies and a flatter IMF slope; $N_{\star}$ and $M_{\rm max}$ are almost unchanged. However, a more sophisticated prescription is necessary to fully explore this scenario. 

\subsubsection{Type Ia SNe}
\indent Type Ia SNe are another potential source of metals in the most metal-poor DLAs. For many decades, it has been appreciated that the combination of [$\alpha$/Fe] and [Fe/H] can indicate when a system has been chemically enriched by SNe Ia (see the discussion by \citealt{Tinsley1979} and \citealt{Wheeler1989}). Type Ia SNe occur after long-lived, low mass stars have become white dwarfs, therefore there is a delay in the onset of chemical enrichment from these SNe compared to that of Type II core-collapse SNe. The yields of Type Ia SNe are rich in Fe-peak elements, while those of Type II SNe are rich in both $\alpha$-capture and Fe-peak elements. The short lifespans of massive stars mean that an environment is first enriched with the products of Type II core-collapse SNe.  This produces an IMF-weighted abundance ratio of [$\alpha$/Fe]. As the system evolves, the pool of high mass progenitors is quickly exhausted, and the [$\alpha$/Fe] ratio plateaus until the onset of enrichment by Type Ia SNe. The Fe-rich ejecta of these SNe cause a decline in [$\alpha$/Fe], known as the `metallicity-knee' (or `$\alpha$-knee'). This can be observed by measuring the abundances of stars over a range of metallicities in a galaxy \citep{Matteucci1990, Matteucci2003}. In the Milky Way, the knee occurs at [Fe/H] $\approx -1$, while for some dwarf spherodial galaxies (dSphs), the knee has been identified at lower metallicities \citep{TolstoyHillTosi2009}. Sculptor and Fornax, two dSphs, show a decline in [$\alpha$/Fe] at [Fe/H] $\approx -1.8$ and [Fe/H] $\approx-1.9$, respectively \citep{Starkenburg2013, Hendricks2014, Hill2018}. A similarly positioned knee has been observed across the DLA population by \cite{CookePettiniJorgenson2015}. They find that [$\alpha$/Fe] begins to fall when [Fe/H] $\gtrsim -2.0$. 
For the systems used in our analysis [Fe/H] $<-2.0$ (see Table~\ref{tab:CO}); this places our DLA sample in the plateau of [$\alpha$/Fe] and suggests that they have likely not yet been significantly contaminated by Type Ia SNe ejecta. 

\subsubsection{Population II core-collapse SNe}
\indent The ejecta of metal-enriched (i.e. Population II) core-collapse SNe are also a potential source of C, O, Si, and Fe, which may pollute the metal-free (Population III) signature in metal-poor DLAs. As Figure~\ref{fig:CO_v_explos_eng} highlights, at the explosion energies recovered by our fiducial model analysis, the relative yields of the most abundant elements are almost independent of the metallicity of the progenitor star. It is therefore difficult to uniquely delineate Population II versus Population III stars using only the most abundant chemical elements. However, it is nevertheless possible to search for several key chemical signatures in the metal-poor DLA population that might tease out the enrichment by Population III stars, including: (1) a very low value of $N_{\star}$ (e.g. $\sim~1-5$) might indicate that only a few massive stars contributed to the enrichment of the metal-poor DLA population; (2) if the first stars formed from an IMF with a slope parameter, $\alpha$, that is different from Salpeter, we might expect to uncover an evolution of the slope parameter at the lowest metallicities; (3) we could measure the relative chemical abundances of elements near the Fe-peak (e.g. [Zn/Fe]; \citealt{Primas2000, Umeda2002, Cayrel2004}), which may provide a more sensitive diagnostic of enrichment by metal-free stars. This may become possible with the next generation of 30\,m class telescopes. At present, given that we only have access to the most abundant metals, we cannot uniquely distinguish between the yields of metal-free and slightly metal-enriched massive stars. Note that this prediction of a low $N_{\star}$ may be negated if the metal-poor DLAs contain the chemical products of multiple minihalos. However, given the relatively large value of $N_{\star}$ recovered by our fiducial model analysis, in addition to an IMF slope parameter that is consistent with investigations of current star formation, it appears likely that some of the metal-poor DLAs in our sample have been enriched by Population II stars.

\begin{figure*}
  \centering
  \includegraphics[width =\textwidth]{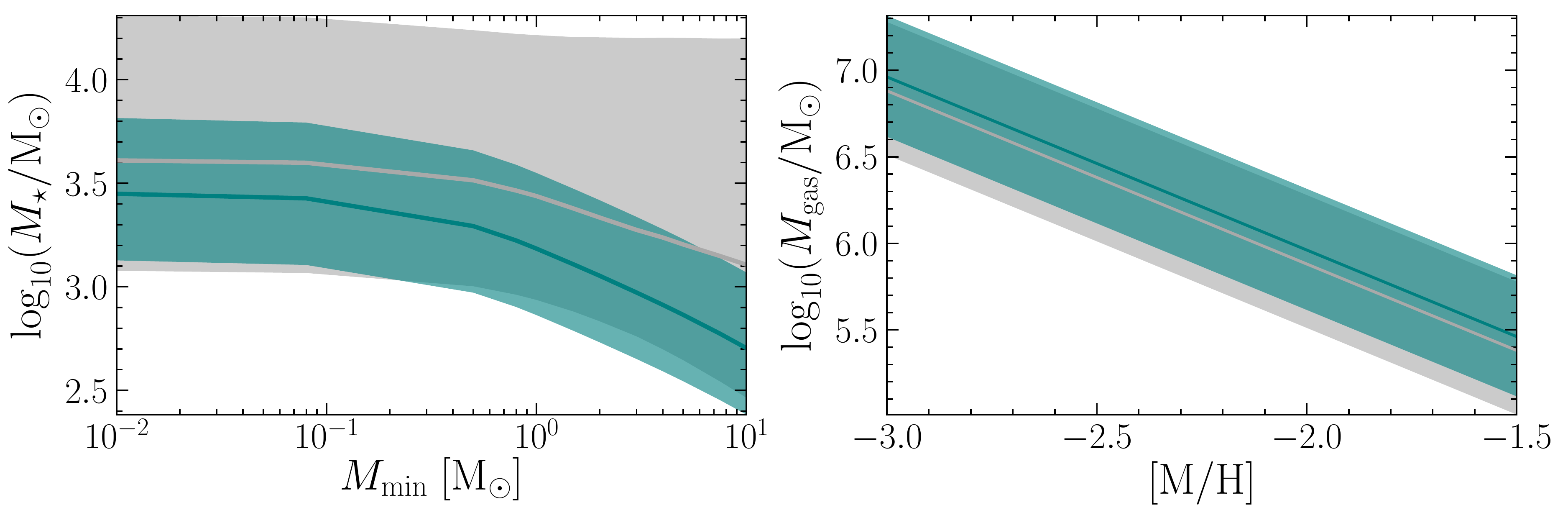}
    \caption{Inferred physical properties of metal-poor DLAs, based on the likelihood distributions of our model parameters. The grey distribution shows the expected stellar content when the model IMF slope is allowed to vary. The green distribution is the resulting stellar content under the assumption of a Salpeter IMF slope. The solid lines indicate the median value and the shaded region encompasses the 16$^{\rm th}$ and 84$^{\rm th}$ percentiles. The left panel shows the total stellar mass within a given system as a function of the minimum mass with which stars can form. In this case we have adopted a \protect\citet{Chabrier2003} IMF and assumed that both low- and high-mass stars have contributed to the total stellar content. The right panel shows the total gas mass expected within these systems as a function of their metal content. The metal content [M/H] has been inferred from that of [O/H], which is a common proxy. Note that for a given position in our model parameter space, there are a range of possible ejected metal masses. In this scenario we assume that the most probable value is representative. We also assume 100 per cent retention of the metals. If some metals were not retained, this would lead us to overestimate the gas mass.}
    \label{fig:DLA_prop}
\end{figure*}

 \subsection{Impact of Yield Choice}
 In this subsection, we consider the impact of our model yield choice. To this end, we have explored several different yield sets to determine the sensitivity of our model parameter inferences to the yields. First, we repeat our analysis considering the SNe yields of massive metal-enriched stars. Specifically, we consider progenitors whose metallicity is $10^{-4}~{\rm Z_{\odot}}$, using the \citetalias{WoosleyWeaver1995} yield calculations. An inspection of the expected abundance ratios under the assumption of the \citetalias{WoosleyWeaver1995} yields, indicates that these yields are less able to reproduce the observed data compared to our fiducial yield choice. We come to the same conclusion when considering the yields of metal-free stars as calculated by \citetalias{LimongiChieffi2012} (see Appendix~\ref{append:corner} for a detailed comparison). We note that the \citetalias{WoosleyWeaver1995} and the \citetalias{LimongiChieffi2012} yields are not calculated across a grid of fixed explosion energies. Across the range of progenitor masses, the final kinetic energy of the SN ejecta varies, but is typically $\sim10^{51}~{\rm erg}$ for both yield sets. To test how this limitation impacts our results, we have repeated our analysis using the \citetalias{HegerWoosley2010} yields, after imposing a strong prior on the SN explosion energy. We found that the model parameter estimates varied significantly between a moderate ($1.2\times10^{51}$ erg) explosion and that of a high energy ($5\times10^{51}$ erg) explosion. Thus being able to include the explosion energy as a free parameter allows the model to find a better fit to the available data. \\
\indent A factor which impacts the yields of these calculations is the adopted rates of both the $3\alpha$ reaction (which creates $^{12}{\rm C}$) and the $^{12}$C($\alpha,\gamma$)$^{16}$O reaction (which destroys $^{12}{\rm C}$). Adopting different determinations of these reaction rates can influence whether $^{12}{\rm C}$ or $^{16}{\rm O}$ is the dominant product of helium burning and, in turn, impact the yields of all elements \citep{WeaverWoosley1993}. Currently, these reaction rates have an associated uncertainty of $\sim10$ per cent \citep{West2013}. \citetalias{HegerWoosley2010} adopt a $^{12}$C($\alpha,\gamma$)$^{16}$O reaction rate comparable to the most recent determination by \cite{An2016} who recommend a reaction rate\footnote{This value corresponds to $S_{\rm tot}(300~{\rm keV})=(167.2\pm 7.3)~{\rm keV~b}$. This value agrees fortuitously well with the rates adopted by \citetalias{HegerWoosley2010} (175~keV b), \citetalias{WoosleyWeaver1995} (170~keV b), and \citetalias{LimongiChieffi2012} (165~keV b).} of $(7.83\pm 0.35)\times 10^{15}\rm\; cm^{3}mol^{-1}s^{-1}$ at T = $9\times10^{8}\rm\; K$, the temperature at which stellar helium burning occurs. Therefore, given the accuracy of the reaction rate adopted by \citetalias{HegerWoosley2010}, in combination with the fine mass resolution and explosion energy grids, and the fact that the \citetalias{HegerWoosley2010} models more accurately reproduce the available data (see Appendix~\ref{append:corner}), we consider the \citetalias{HegerWoosley2010} yields to be the superior choice for our analysis. \\

\subsection{Inferred properties of DLAs}
Given the fiducial results of our enrichment model analysis, we now investigate some of the typical physical properties of the DLAs in our sample. These systems are only seen in absorption. Directly determining their total stellar content would be challenging, and we have no direct means to observationally investigate their total gas content. However, our analysis provides an indication of the enriching stellar population, which can be used to extrapolate an estimate of the total stellar mass and gas mass\footnote{Recall, at the explosion energies recovered by our fiducial model analysis, the relative yields of both [C/O] and [Si/O], for a given progenitor mass, appear to be almost indistinguishable between Population II and Population III stars (see Figure~\ref{fig:CO_v_explos_eng}). Therefore, in the following calculations we consider the \citetalias{HegerWoosley2010} yields to be an appropriate estimator of both Population II and Population III core-collapse SNe yields.}. \\
\indent In what follows, we use the parameter distributions of our enrichment model analysis to describe the IMF of the enriching population, $\xi(M)$. 
The total stellar mass of a typical system can then be inferred using the equation:
\begin{equation}
    M_{\star} = \int_{M_{\rm min}}^{M_{\rm max}} \xi(M)M{\rm d}M~,
\end{equation}
where $\xi(M)$ represents the IMF of the system. Note that our enrichment model is only sensitive to the yields of stars $>10~{\rm M_{\odot}}$. 
Therefore, if we assume that low mass stars have also formed in very metal-poor DLAs (as would be expected if these gas clouds have been enriched by Population II stars), these stars will constitute a significant fraction of the total stellar mass. For this inference, we must consider an IMF that is best able to account for the contribution of both low mass and high mass stars. In what follows, we adopt a \cite{Chabrier2003} IMF, such that stars below $1~{\rm M_{\odot}}$ are modelled by a log-normal distribution.
Given a bottom-heavy IMF, stars more massive than $100~{\rm M_{\odot}}$ provide a negligible contribution to the total stellar mass of a system, therefore we adopt an upper mass limit of $M_{\rm max} = 100{\rm M_{\odot}}$. The left panel of Figure~\ref{fig:DLA_prop} shows the total stellar mass inferred for a typical metal-poor DLA as a function of the minimum mass with which stars can form. We show the inferred stellar mass from both our fiducial model analysis and the case of a Salpeter IMF slope at high masses. From this we see that, for the case of a Salpeter IMF slope, if the stars within very metal-poor DLAs can form down to $0.1~{\rm M_{\odot}}$, then the total stellar mass formed over the lifetime of the system is $\log_{10}(M_{\star})=3.5^{+0.3}_{-0.4}~{\rm M_{\odot}}$. This value is comparable to the stellar content of the faint Milky Way satellite population \citep{MartinDeJongRix2008, McConnachie2012}, which typically span a mass range of $\sim(10^{2}-10^{5})~{\rm M_{\odot}}$, and are still expected to contain gas at redshift $z\sim3$ \citep{Onorbe2015, Wheeler2018}. We suggest that, given a more robustly determined enrichment model, our inference could allow us to draw parallels between the metal-poor DLA population and their potential galactic descendants. A precise inference of the stellar mass content could also help discern whether the most metal-poor DLAs are the progenitors of some Ultra Faint Dwarf (UFD) galaxies.
 \\
\indent We can also use the results of our analysis to infer the typical total gas mass of metal-poor DLAs. Given an enrichment model, we can determine the mass of metals that have been introduced, to a previously pristine environment, through SNe ejecta. In the same way that we construct the distribution of [C/O] for a given enrichment model (described in Section~\ref{sec:methods}), we can also construct a distribution of the ejected metal mass. For simplicity, we take the most probable value of this distribution to be representative. Thus, by sampling the parameter distributions shown in Figure~\ref{fig:corner_fid} and calculating the associated ejected metal mass, we can build a distribution of the typical metal mass expected within our systems. We can then infer the typical mass of gas that has been mixed with the metals of core-collapse SNe, as a function of the measured [O/H] metallicity of the gas. For a given [O/H] abundance and ejected oxygen mass, Equation~\ref{eqn:X/Y} can be used to determine the expected mass of hydrogen that has been mixed with the metals of the Type II core-collapse SNe yields. As metals contribute a negligible amount to the overall system mass, the total gas mass is given by the sum of the contribution from both hydrogen and helium. We assume that the helium mass fraction is equal to the primordial value ($Y_{\rm P}\simeq 0.247$; \citealt{Pitrou2018}). The right panel of Figure~\ref{fig:DLA_prop} shows the total gas mass of a typical system as a function of the system's metal abundance. For an extremely metal-deficient system i.e. [M/H] $\sim -3.0$, the total gas mass, under the assumption of a Salpeter IMF slope for high mass stars, is $\log_{10}({M_{\rm gas}}/~{\rm M_{\odot}})= 7.0^{+0.3}_{-0.4}$. This suggests that stars may constitute just $\approx~0.03$~per cent of the mass fraction of the most metal-deficient DLAs.  As a point of comparison,  \citealt{CookePettiniJorgenson2015} found that the mass of warm neutral gas within a typical metal-poor DLA is $\log_{10}(M_{\rm WNM}/~{\rm M_{\odot}})=5.4^{+1.9}_{-0.9}$. This was calculated using a sample of DLAs with a typical [O/H] abundance of [O/H]~$\approx -2.0$. Our calculation of the total gas mass within systems of this metallicity suggests that warm neutral gas may constitute $\approx$ 30 per cent of the total gas mass.

\section{Conclusions}
\label{sec:conclusions}
We present a novel, stochastic chemical enrichment model to investigate Population III enriched metal-poor DLAs using their relative metal abundances; this model considers the mass distribution of the enriching stellar population, the typical explosion energy of their SNe, and the average number of enriching stars. We use this model to investigate the chemical enrichment of the 11 most metal-poor DLAs at $z\geq 2.6$. We conduct a maximum likelihood analysis of the enrichment model parameters, given relative abundances ([C/O], [Si/O] and [Fe/O]) of this sample of metal-poor DLAs. Our main conclusions are as follows:
\begin{enumerate}
    \item The mass distribution of the stars that have enriched the sample of metal-poor DLAs can be well-described by a Salpeter-like IMF slope.
    \item The average system has been enriched by $\lesssim72$ massive stars (95 per cent confidence), with a maximum likelihood value of $\hat{N}_{\star}=10\pm 4$, suggesting that the most metal-poor DLAs are minimally enriched. 
    \item Our maximum likelihood estimate of the upper mass limit of enriching stars suggests that the most metal-poor DLAs have been predominantly enriched by stars with masses $\lesssim40~{\rm M_{\odot}}$. This provides tentative evidence in support of the suggestion that some stars above $20~{\rm M_{\odot}}$ fail to explode, and instead collapse directly to black holes \citep{Sukhbold2016}.
    \item Our model suggests that the stars that enriched the most metal-poor DLAs had a typical explosion energy $E_{\rm exp}=1.8^{+0.3}_{-0.2}\times10^{51}~{\rm erg}$, which is somewhat lower than that found by recent works that model the enrichment of metal-poor halo stars \citep{Ishigaki2018, GrimmettHeger2018}. 
    \item Using the results of our likelihood analysis, we infer some of the typical physical properties of metal-poor DLAs. We find that the total stellar mass content of metal-poor DLAs is $\log_{10}(M_{\star}/{\rm M_{\odot}})=3.5^{+0.3}_{-0.4}$, assuming a \cite{Chabrier2003} IMF. We note that this value is comparable to the stellar mass content of faint Milky Way satellites \citep{MartinDeJongRix2008, McConnachie2012} and suggest that, in the future, this inference might allow us to test if some of the most metal-poor DLAs are the antecedents of the UFD galaxy population.
    \item We also infer the total gas mass of typical metal-poor DLAs as a function of their measured [O/H] metallicity: $\log_{10}({M_{\rm gas}}/~{\rm M_{\odot}})= 7.0^{+0.3}_{-0.4}$ for DLAs with [O/H]~$\approx -3.0$. Comparing this value to the mass of warm neutral gas in metal-poor DLAs \big($\log_{10}(M_{\rm WNM}/~{\rm M_{\odot}})=5.4^{+1.9}_{-0.9}$; \citealt{CookePettiniJorgenson2015}\big), we find that $\approx30$ per cent of the gas in a DLA with [O/H]~$\approx-2.0$ may be in the warm neutral phase.
\end{enumerate}
 Finally, we realise the potential for future improvement if we can minimise the potential for contamination from later generations of star formation. Once there is a larger sample of EMP DLAs, we will be able to restrict our analysis to systems with [O/H]$\leq-3.0$. Alternatively, in the future, we will include in our enrichment model the potential contribution of metals from Population II stars (i.e. by considering the mass loss from intermediate mass AGB stars). \\
\indent We conclude by suggesting that our stochastic enrichment model, combined with the \citetalias{HegerWoosley2010} nucleosynthetic yields, is a powerful tool to learn about the earliest episodes of star formation. We expect that future applications of this analysis will reveal a distinctive Population III signature and the opportunity to learn about the mass distribution of the first stars; to this end, we will use our model to explore the enrichment of the most metal-poor stars found in the halo of the Milky Way. Through these systems, we hope to gauge the multiplicity of the first generation of stars.

\section*{Acknowledgements}
We thank an anonymous referee who provided a prompt and careful review of our paper. We thank A. Heger for helpful discussions about the HW10 yield calculations.
During this work, R.~J.~C. was supported by a
Royal Society University Research Fellowship.
We acknowledge support from STFC (ST/L00075X/1, ST/P000541/1).
This project has received funding from the European Research Council 
(ERC) under the European Union's Horizon 2020 research and innovation 
programme (grant agreement No 757535).
This work used the DiRAC Data Centric system at Durham University,
operated by the Institute for Computational Cosmology on behalf of the
STFC DiRAC HPC Facility (www.dirac.ac.uk). This equipment was funded
by BIS National E-infrastructure capital grant ST/K00042X/1, STFC capital
grant ST/H008519/1, and STFC DiRAC Operations grant ST/K003267/1
and Durham University. DiRAC is part of the National E-Infrastructure.
This research has made use of NASA's Astrophysics Data System.





\bibliographystyle{mnras}
\bibliography{references} 


\appendix
\section{Further models}
\label{append:corner}
\begin{figure*}
\centering
    \includegraphics[width =\textwidth]{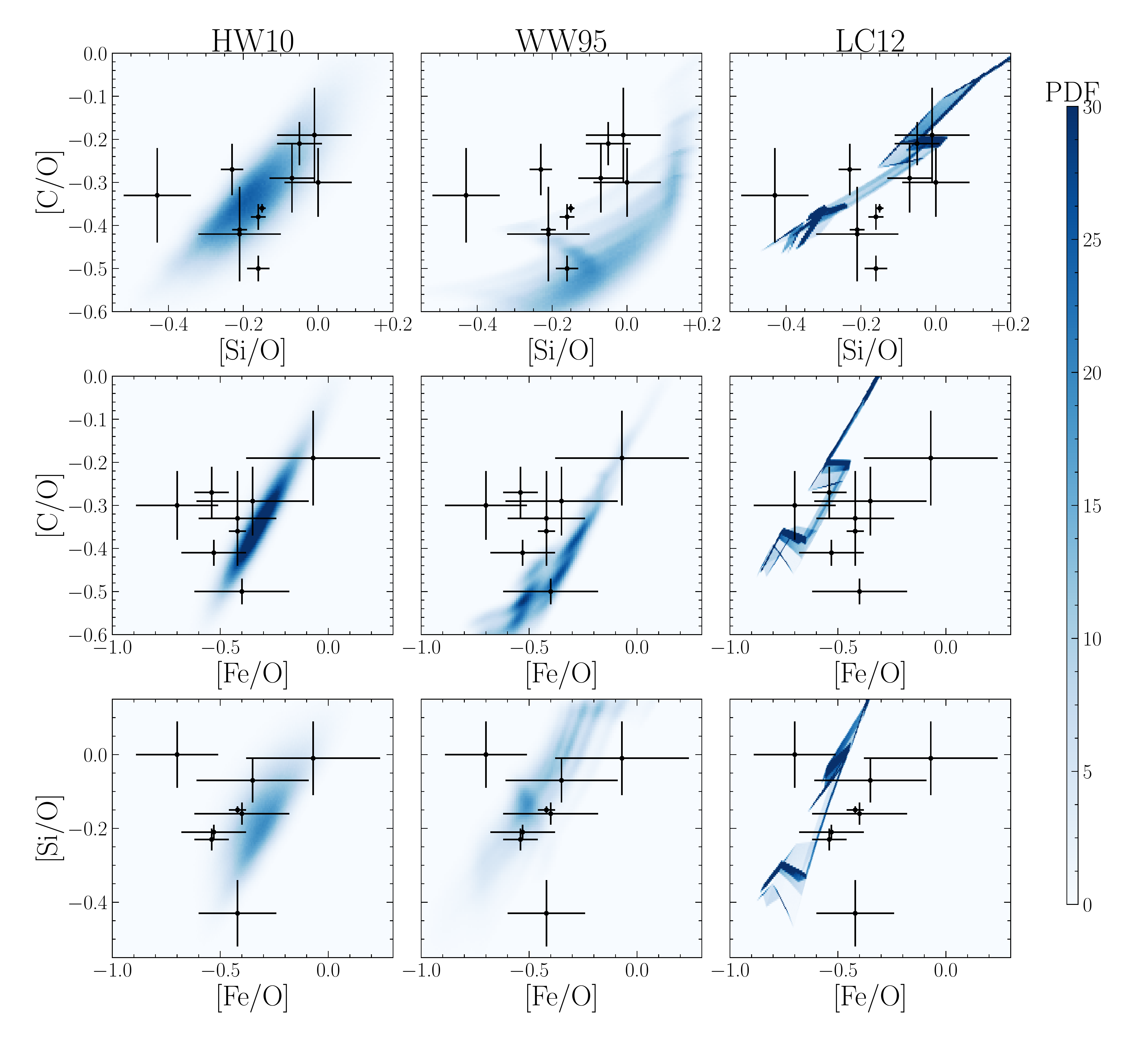}
    \caption{The observed abundance ratios of the most metal-poor DLAs (black symbols with errors) are overplotted on the maximum likelihood parameter distribution ($p(R_{i} \vert R_{m})$; blue background PDF) based on three yield sets. Each column corresponds to a different yield set. From left to right, the underlying yields correspond to: \protect\citetalias{HegerWoosley2010} (fiducial yield choice), \protect\citetalias{WoosleyWeaver1995}, and \protect\citetalias{ LimongiChieffi2012}. Each panel showcases the joint probability density of two expected abundance ratios, given the maximum likelihood enrichment model parameters for a given yield set. The combined inspection of each column gives an indication of $p (R_{i} | R_{m})$, and the ability of a given yield set to simultaneously reproduce the [C/O], [Si/O], and [Fe/O] abundance ratios observed within the metal-poor DLA population.}
    \label{fig:model_fits}
\end{figure*}

In this Appendix, we explore the sensitivity of our model parameter inferences to the adopted nucleosynthesis yield calculation (see also, Section~\ref{sec:disc}). We first consider the yields of massive Population II ($Z=10^{-4}~{\rm Z}_{\odot}$) stars reported by \citetalias{WoosleyWeaver1995}. These yields have been calculated for a typical explosion energy of $1.2\times10^{51}~{\rm erg}$. We therefore only consider three model parameters: $\alpha$, $N_{\star}$, and $M_{\rm max}$. The maximum mass considered by the \citetalias{WoosleyWeaver1995} yields is $M_{\rm max}=40~{\rm M_{\odot}}$. We repeat the analysis described in Section~\ref{sec:fiducial} to find the enrichment model that best describes the abundance ratios observed in the most metal-poor DLAs. 
We also repeat our analysis considering the model yield calculations of massive metal-free stars reported by \citetalias{LimongiChieffi2012}. These yields have been calculated for a typical explosion energy of $\sim 10^{51}~{\rm erg}$. In Figure~\ref{fig:model_fits}, we show the maximum likelihood enrichment models (blue PDF) based on each of the above yield sets, and compare these to the observed data. From this we can see that the enrichment model indicated by the \citetalias{HegerWoosley2010} yields produces the best overall fit to the observed data. This, alongside the fine mass resolution and the detailed consideration of the explosion energy afforded by the \citetalias{HegerWoosley2010} yields, reaffirms our choice to use this yield set in our fiducial analysis. 

\bsp    
\label{lastpage}
\end{document}